\documentclass[twocolumn,a4paper,10pt]{article}
\usepackage{geometry}
\usepackage{amsmath,amsfonts}
\usepackage{epsfig}
\usepackage{times}
\usepackage{epsfig}
\usepackage[small,hang]{caption2}

\usepackage{graphicx} % Required for the inclusion of images
\usepackage[authoryear,sort&compress]{natbib} % Required to change bibliography style to APA
\setcitestyle{authoryear,open={(},close={)}}
\usepackage{xcolor}
\usepackage{soul}
\usepackage{float}
\usepackage{hyperref}
\usepackage[capitalise]{cleveref}
\usepackage[percent]{overpic}
\makeatletter

\newcounter{numbersec}
\renewcommand{\section}[1]{\par\noindent\stepcounter{numbersec}
\par
\vspace{6pt}
\noindent\textbf{\large   \arabic{numbersec} \hspace*{0.3cm} #1 }
\par
\vspace{2pt}
}
\renewcommand{\subsection}[1]{
\par
\vspace{6pt}
\noindent\textbf{#1}
\par
}
\renewcommand{\subsubsection}[1]{%
\par
\vspace{6pt}
\textbf{#1.}
}

\newcommand{\Abstract}{\par\vspace{6pt}\noindent\textbf{\large Abstract}\par\vspace{2pt}}
\newcommand{\Acknowledgments}{\par\vspace{6pt}\noindent\textbf{\large Acknowledgments }\par\vspace{2pt}}

\makeatother

\title{\vspace*{-12mm}
\LARGE \sc \textbf{
Physics-informed neural networks for solving Reynolds-averaged Navier--Stokes equations
}}
\author{ \Large \bf \textit{ 
    Hamidreza Eivazi$^{1,2}$, Mojtaba Tahani$^{1}$, Philipp Schlatter$^{2}$, Ricardo Vinuesa$^{2}$}  \\ \\
\bf  $^{1}$ \textit{Faculty of New Sciences and Technologies, University of Tehran, Tehran, Iran} \\
\bf  $^{2}$ \textit{SimEx/FLOW, Engineering Mechanics, KTH Royal Institute of Technology,} \\
\bf  \textit{SE-100 44 Stockholm, Sweden} \\
\underline{\bf \href{mailto:hamidre@kth.se}{hamidre@kth.se}}
}
\date{}

\begin{document}
%%%%%%%%%%%%%%%%%%%%%%%%%%%%%%%%%%%%%%%%%%%%%%%%%%%%%%%%%%%%%%%%%%%%%%%%%%%%%%%%%%%%%%%%%%%%
%
%\multicolumn{2}{c}{ffffffffffffffffffffffffffffffffffffffff}

%\begin{multicols}{2}[\section{Titre numgtggggggggggggggggggggggggggggggggggggggggéroté.}]
%   blabla sur deux colonnes, c'est plus sérieux. C'est le
%   style qui est généralement utilisé pour écrire des
%   articles.
%\end{multicols}

%
%\begin{multicols}{1}
%   ffffffffffffffffffffffffffffffffffffffffffffffffffffffffffffffffffffffffffffffffffffffffff
%\end{multicols}

\maketitle
\thispagestyle{empty}

%\multicolumn{2}{\centering}{textddddddddddddddddde}

%\Title{Reconstruction of turbulent fluctuations for hybrid RANS/LES simulations using a synthetic eddy method}
%********** Insert here the name(s) and address(es) of the author(s).
%\Author{A.~Author, A.~N.~Otherauthor}
%\Author{N. Jarrin, R. Prosser, F. Billard and D. Laurence}
%\Address{School of Mechanical, Aerospace and Civil Engineering, }
%\Address{The University of Manchester, Manchester M60 1QD, UK}
%%\Address{$^+$ EDF R\&D, 6 quai Watier, 78420 Chatou, France}
%\Email{ N.Jarrin@postgrad.manchester.ac.uk,  }
%%\Email{ r.prosser@manchester.ac.uk, dominique.laurence@manchester.ac.uk}

%
%%%%%%%%%%%% Insert here the abstract body.
%
\Abstract

Physics-informed neural networks (PINNs) are successful machine-learning methods for the solution and identification of partial differential equations (PDEs). We employ PINNs for solving the Reynolds-averaged Navier--Stokes (RANS) equations for incompressible turbulent flows without any specific model or assumption for turbulence, and by taking only the data on the domain boundaries. We first show the applicability of PINNs for solving the Navier--Stokes equations for laminar flows by solving the Falkner–Skan boundary layer. We then apply PINNs for the simulation of four turbulent-flow cases, \emph{i.e.}, zero-pressure-gradient boundary layer, adverse-pressure-gradient boundary layer, and turbulent flows over a NACA4412 airfoil and the periodic hill. Our results show the excellent applicability of PINNs for laminar flows with strong pressure gradients, where predictions with less than 1\% error can be obtained. For turbulent flows, we also obtain very good accuracy on simulation results even for the Reynolds-stress components.

%for the supervised learning. In the RANS equations, the number of unknowns is larger than the number of equations---this is the so-called closure problem. Traditional solvers require a model for turbulence to close the system of equations. We solve this problem through the use of the data on the domain boundaries (including Reynolds-stress components) along with the RANS equations that guide the learning process towards the solution. 
%
%%%%%%%%%%%% Body of the article.
%
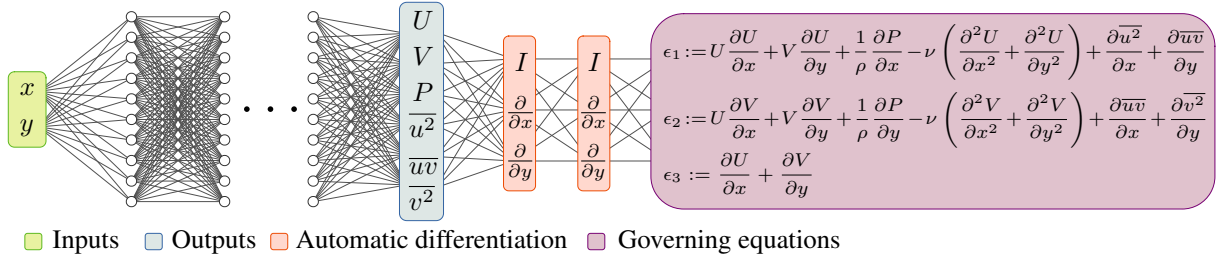
\begin{figure*}[th]
    \centering
    \begin{overpic}[width=1\textwidth]{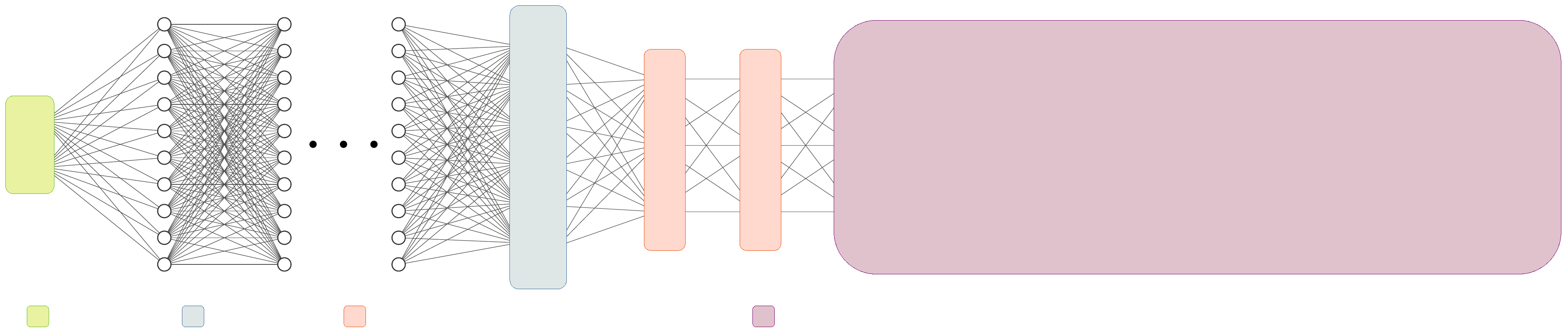}
    \put (1.3, 13) {$x$}
    \put (1.3, 10) {$y$}
    \put (33.6, 18.5) {$U$}
    \put (33.6, 15.5) {$V$}
    \put (33.6, 12.5) {$P$}
    \put (33.3, 9.5) {$\overline{u^2}$}
    \put (33.3, 6.5) {$\overline{uv}$}
    \put (33.3, 3.5) {$\overline{v^2}$}
    \put (42, 15) {$I$}
    \put (41.2, 11) {$\frac{\partial}{\partial x}$}
    \put (41.2, 7) {$\frac{\partial}{\partial y}$}
    \put (48, 15) {$I$}
    \put (47.2, 11) {$\frac{\partial}{\partial x}$}
    \put (47.2, 7) {$\frac{\partial}{\partial y}$}
    \put (53.8, 16.2) {\scriptsize{
    $\epsilon_1 \!:=\! U\dfrac{\partial U}{\partial x} \!+\! V\dfrac{\partial U}{\partial y} \!+\! \dfrac{1}{\rho} \dfrac{\partial P}{\partial x} \!-\! \nu \left (\dfrac{\partial^2 U}{\partial x^2} \!+\! \dfrac{\partial^2 U}{\partial y^2} \right ) \!+\! \dfrac{\partial \overline{u^{2}}}{\partial x} \!+\! \dfrac{\partial \overline{uv}}{\partial y}$
    }}
    \put (54.2, 10.6) {\scriptsize{$
    \epsilon_2 \!:=\! U\dfrac{\partial {V}}{\partial x} \!+\! {V}\dfrac{\partial {V}}{\partial y} \!+\! \dfrac{1}{\rho} \dfrac{\partial {P}}{\partial y} \!-\!  \nu \left (\dfrac{\partial^2 {V}}{\partial x^2} \!+\! \dfrac{\partial^2 {V}}{\partial y^2} \right) \!+\! \dfrac{\partial \overline{uv}}{\partial x} \!+\! \dfrac{\partial \overline{v^{2}}}{\partial y}
    $}}
    \put (54.2, 6) {\scriptsize{$
    \epsilon_3 := \dfrac{\partial {U}}{\partial x} + \dfrac{\partial {V}}{\partial y}
    $}}
    \put (4, 0.3) {Inputs}
    \put (13.8, 0.3) {Outputs}
    \put (24, 0.3) {Automatic differentiation}
    \put (50.5, 0.3) {Governing equations}
    
    \end{overpic}
    \caption{A schematic of PINNs for solving the RANS equations for a general two-dimensional set-up. The left part of the model is a FNN, and the right part is the formulation of the RANS equations using AD.}
    \label{fig1}
\end{figure*}

\section{Introduction}

In recent years, machine-learning (ML) methods have started to play a revolutionary role in many scientific disciplines \citep{Vinuesa2020}. Fluid mechanics has been one of the active research topics for development of innovative ML-based approaches \citep{kutz_2017,duraisamy_et_al,brunton_et_al_2020}. The contribution of ML in turbulent-flow problems is mainly in the contexts of data-driven turbulence closure modeling \citep{ling2016,Jiang_2021}, prediction of temporal dynamics in turbulent flows \citep{srinivasan,EIVAZI_2021}, nonlinear modal decomposition \citep{murata_fukami_fukagata_2020,Eivazi_2020} extraction of turbulence theory form data \citep{jimenez_2018}, non-intrusive sensing in turbulent flows \citep{guastoni_2020,guemes2021}, and flow control \citep{RL}. More recently, exploiting the universal-approximation property of neural networks for solving complex partial differential equation (PDE) systems has brought attention, aiming to provide efficient solvers that approximate the solution. Physics-informed neural networks (PINNs), introduced by \cite{raissi_pinns}, have been shown to be well suited for the solution of forward and inverse problems related to several different types of PDEs. PINNs have been used to simulate vortex-induced vibrations \citep{raissi_viv} and to tackle ill-posed inverse fluid mechanics problems \citep{raissi_sci}. Moreover, PINNs have been employed for super-resolution and denoising of 4D-flow magnetic resonance imaging (MRI) \citep{FATHI2020} and prediction of near-wall blood flow from sparse data \citep{arzani2021}. Recently, \cite{NSFnets} showed the applicability of PINNs for the simulation of turbulence directly, where good agreement was obtained between the direct numerical simulation (DNS) results and the PINNs simulation results. A detailed discussion on the prevailing trends in embedding physics into ML algorithms and diverse applications of PINNs can be found in the work by \cite{Karniadakis2021}.

In this study, we employ PINNs for solving the Reynolds-averaged Navier--Stokes (RANS) equations for incompressible turbulent flows without any specific model or assumption for turbulence. In the RANS equations, the loss of information in the averaging process leads to an underdetermined system of equations. Traditional solvers require the introduction of modeling assumptions to close the system. We introduced an alternative approach to tackle this problem by using the information from a few data examples and the underdetermined system of equations for the training of a neural network that solves the system of equations. In particular, we use the data on the domain boundaries (including Reynolds-stress components) along with the RANS equations that guide the learning process towards the solution. The spatial coordinates are the inputs of the model and the mean-flow quantities, \emph{i.e.}, velocity components, pressure, and Reynolds-stress components, are the outputs. Automatic differentiation (AD) \citep{baydin2018} is applied to differentiate the outputs with respect to the inputs to construct the RANS equations. Only the data on the domain boundaries are considered as the training dataset for the supervised-learning part while a set of points inside the domain together with the points on the boundaries are used to calculate the residual of the governing equations, which acts as the unsupervised loss. The Reynolds number is set through the governing equations. We take mean-flow quantities obtained from DNS or well-resolved large-eddy simulation (LES) of canonical turbulent flow cases as the reference.

% \begin{figure*}[th]
%     \centering
%     \begin{overpic}[width=0.96\textwidth]{PINNs_RANS_Schematic.pdf}
%         \put (-2,20) {(a)}
%     \end{overpic}

%     \vspace{0.5cm}

%     \begin{overpic}[width=0.96\textwidth]{ZPG_ETMM.pdf}
%         \put (-2,43) {(b)}
%         \put (49,43) {(c)}
%         \put (74,43) {(d)}
%     \end{overpic}
%     \caption{A schematic of PINNs for solving the RANS equations, and the simulation results of ZPG boundary layer using PINNs in comparison with the reference data. (a) PINN model for a general two-dimensional set-up. The left part of the model is a FNN, and the right part is the formulation of the RANS equations using AD. (b) Contours of $U$ (top), $V$ (middle), and $\overline{uv}$ (bottom); (b) Shape factor $H_{12}$ (top) and skin-friction coefficient $c_f$ (bottom). (d)  Inner-scaled mean streamwise velocity $U^+$(top) and Reynolds-stress $\overline{uv}^+$ (bottom) profiles at $Re_{\theta} = 2500$, 4000, and 5500. Dashed and solid lines represent, respectively, the PINN predictions and the reference data. Darker color shows higher $Re_{\theta}$.}
%     \label{fig1}
% \end{figure*}

\section{Methodology}

A schematic of PINNs for the RANS equations is depicted in \Cref{fig1}. In a general two-dimensional set-up, the spatial coordinates ($x$ and $y$) are the inputs of a fully-connected neural network (FNN), and the outputs are the mean streamwise and wall-normal components of velocity ($U$ and $V$, respectively), pressure ($P$), and Reynolds-stress components ($\overline{u^{2}}$, $\overline{uv}$, and $\overline{v^{2}}$). Automatic differentiation \citep{baydin2018} is applied to differentiate outputs with respect to the inputs and formulate the RANS equations (continuity and momentum equations). AD can be implemented directly from the deep learning framework as it is used to compute the gradients and update the network parameters, i.e. weights $\boldsymbol{w}$ and biases $\boldsymbol{b}$, during the training. We use the open-source machine-learning software framework TensorFlow \citep{tensor_flow} to develop our PINN models. TensorFlow provides the ``\texttt{tf.GradientTape}" application programming interface (API) for AD by computing the gradient of a computation with respect to some inputs. TensorFlow ``records" the computational operations executed inside the context of a \texttt{tf.GradientTape} onto a so-called ``tape", and then uses that tape to compute the gradients of a recorded computation using reverse-mode differentiation.

In our setup, only the data on the domain boundaries are used as the training dataset for supervised learning. The total loss is the summation of the supervised loss and the residual of the governing equations as follows:
\begin{subequations}
    \begin{equation}
        L = L_e + L_b,
    \end{equation}
    \begin{equation}
        L_e = \dfrac{1}{N_e} \sum_{i=1}^3 \sum_{n=1}^{N_e} |\epsilon_i^n|^2,
    \end{equation}
    \begin{equation}
        L_b = \dfrac{1}{N_b} \sum_{n=1}^{N_b} |\mathbf{U}_b^n - \tilde{\mathbf{U}}_b^n|^2,
    \end{equation}
\end{subequations}
where $L_e$ and $L_b$ are the loss-function components corresponding to the residual of the RANS equations and the boundary conditions, respectively. Here $N_e$ represents the number of points for which the residual of the RANS equations is calculated, and $N_b$ is the number of training samples on the domain boundaries. We consider a set of points inside the domain and compute the residuals on these points together with the points on the domain boundaries. Here, $\mathbf{U}_b^n = [U_b^n, V_b^n, P_b^n, \overline{u^{2}}_b^n, \overline{uv}_b^n, \overline{v^{2}}_b^n]^\mathsf{T}$ represents the given data for point $n$ on the boundaries. $\tilde{\mathbf{U}}_b^n$ is the vector of PINN predictions at the corresponding point, and $\epsilon_i^n$ is the residual of the $i^{th}$ governing equation at point $n$. It is also possible to consider weighting coefficients to balance different terms of the loss function and accelerate convergence in the training process \citep{NSFnets}.

\begin{figure}[th]
    \centering
    \begin{overpic}[width=0.9\columnwidth]{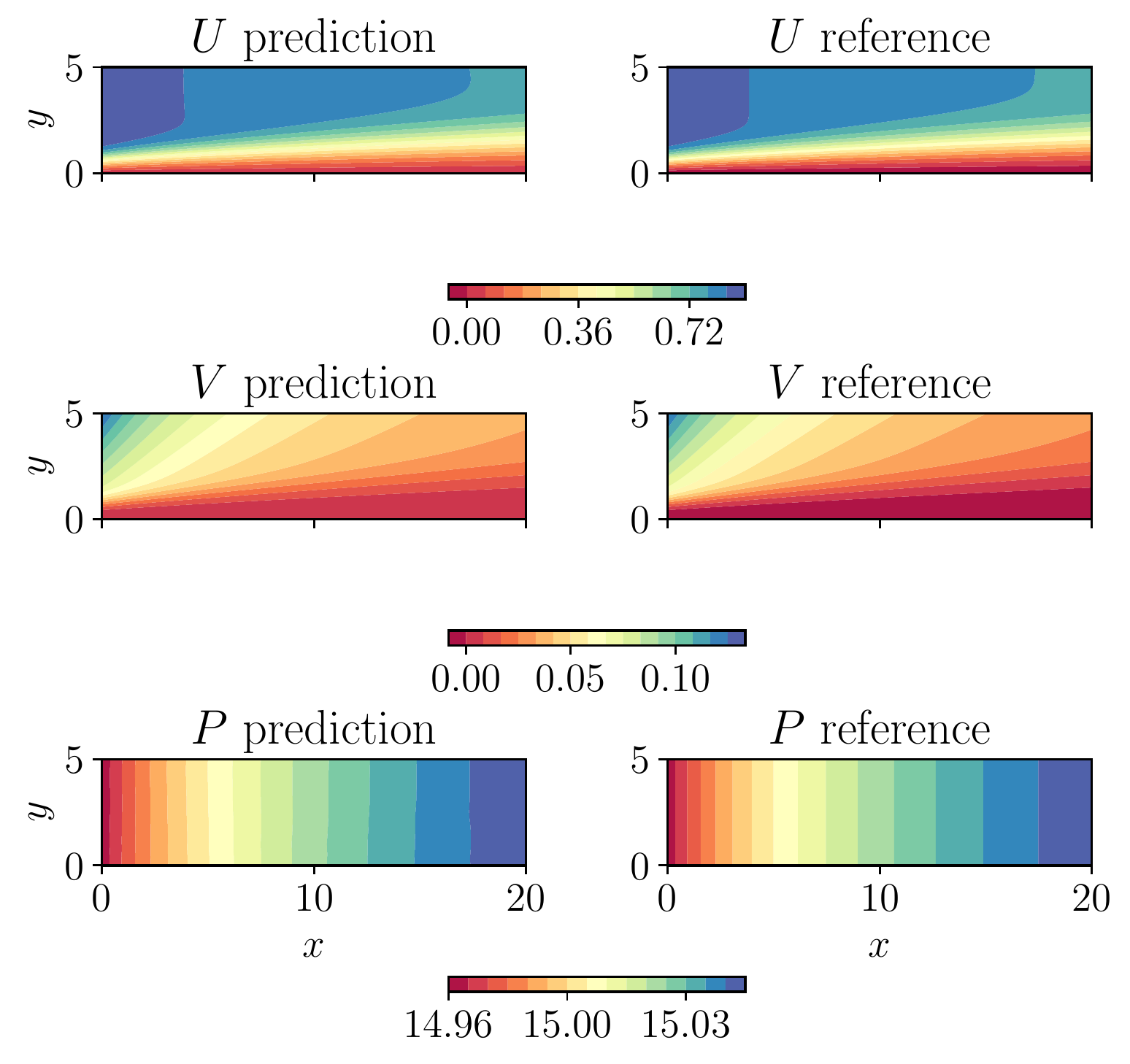}
        \put (-2,90) {(a)}
        \put (-2,60) {(b)}
        \put (-2,30) {(c)}
    \end{overpic}
    \caption{Simulation results of the Falkner--Skan boundary layer with adverse pressure gradient using PINNs (left) in comparison with the reference data (right). (a, b, c) Contours of $U$, $V$, and $P$, respectively.}
    \label{fig2}
\end{figure}

\begin{figure*}[th]
    \begin{overpic}[width=0.96\textwidth]{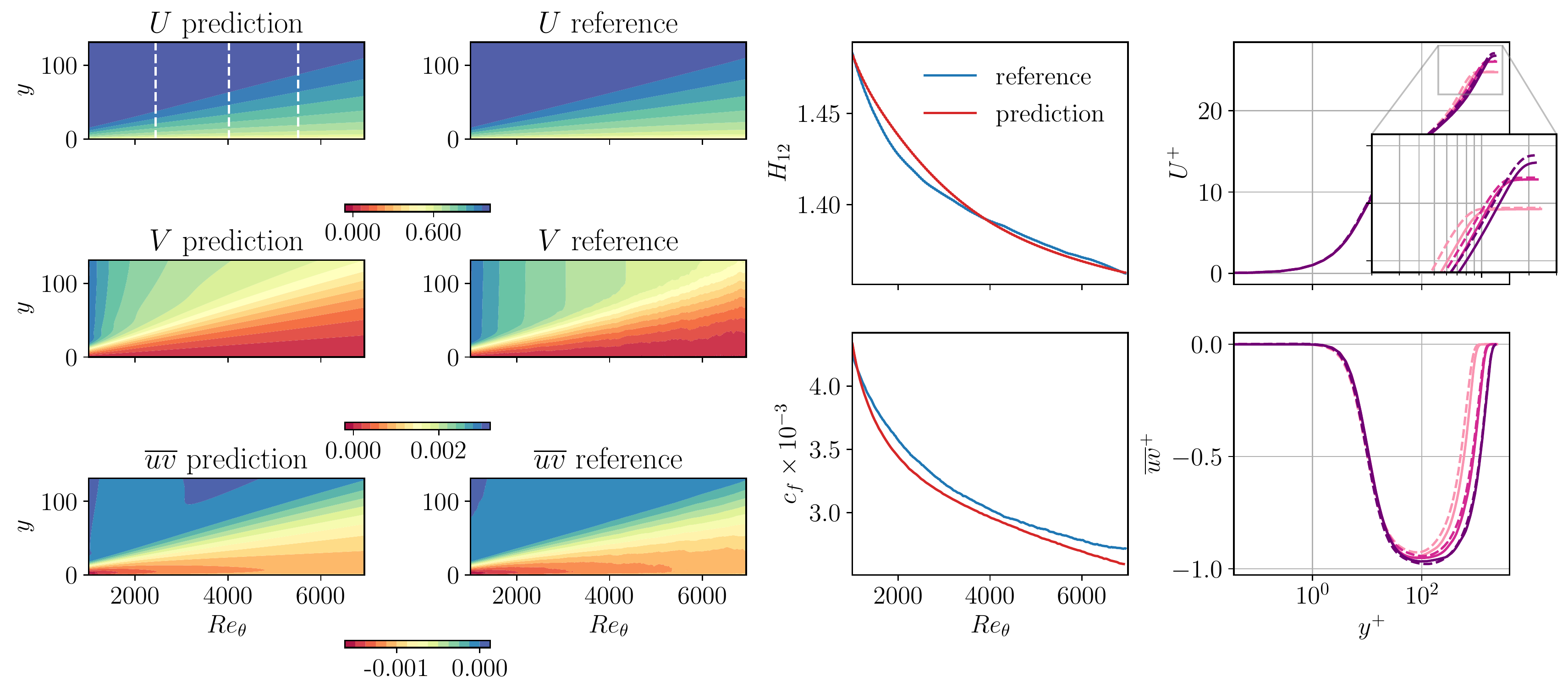}
        \put (-2,43) {(a)}
        \put (49,43) {(b)}
        \put (74,43) {(c)}
    \end{overpic}
    \caption{Simulation results of the ZPG turbulent boundary layer using PINNs in comparison with the reference data. (a) Contours of $U$ (top), $V$ (middle), and $\overline{uv}$ (bottom); (b) Shape factor $H_{12}$ (top) and skin-friction coefficient $c_f$ (bottom). (c)  Inner-scaled mean streamwise velocity $U^+$(top) and Reynolds-stress $\overline{uv}^+$ (bottom) profiles at $Re_{\theta} = 2500$, 4000, and 5500. Dashed and solid lines represent, respectively, the PINN predictions and the reference data. Darker color shows higher $Re_{\theta}$.}
    \label{fig3}
\end{figure*}

\begin{figure*}[th]
    \centering
    \begin{overpic}[width=1\textwidth]{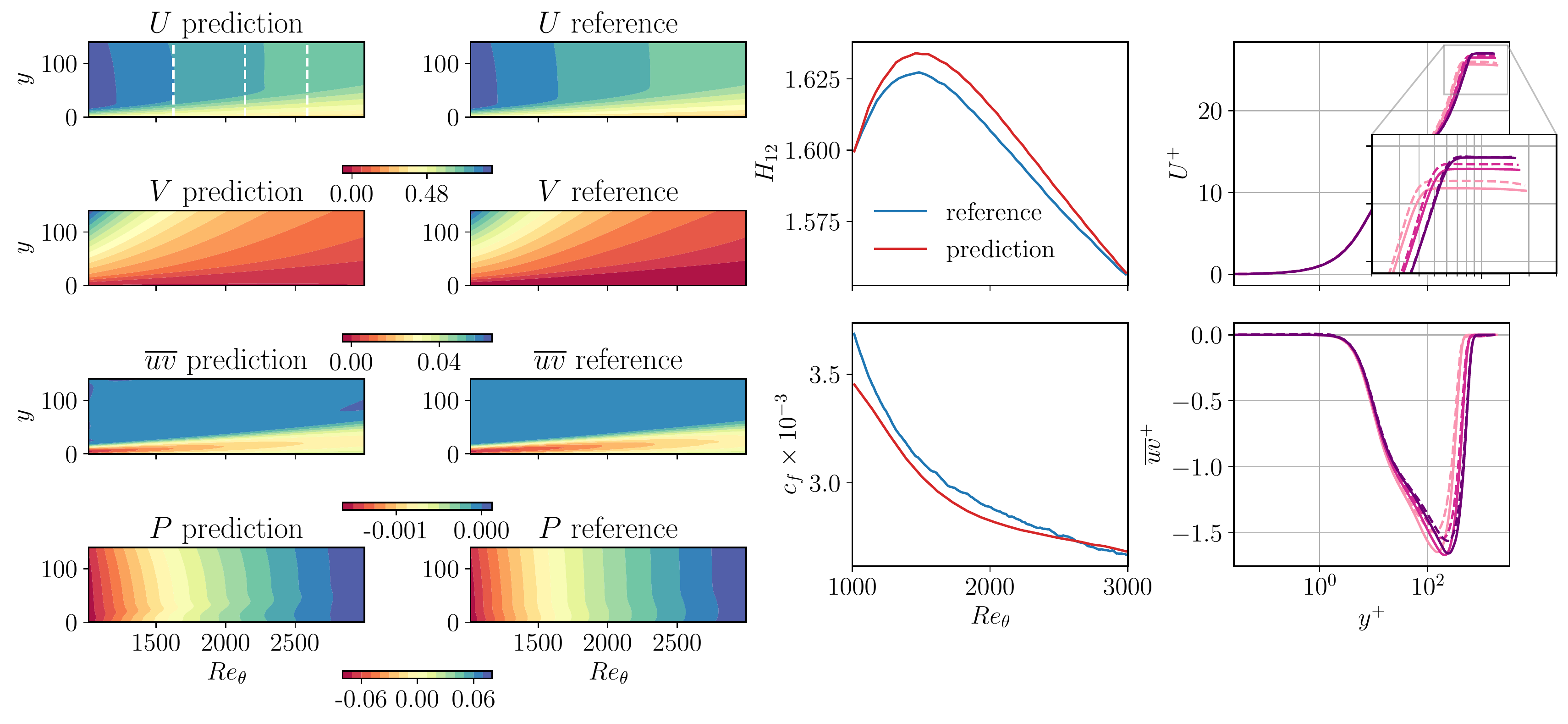}
        \put (-2,45) {(a)}
        \put (49,45) {(b)}
        \put (74,45) {(c)}
    \end{overpic}
    \caption{Simulation results of the APG turbulent boundary layer using PINNs in comparison with the reference data. (a) Contours of $U$, $V$, $\overline{uv}$, and $P$ from top to bottom, respectively. (b) Shape factor $H_{12}$ (top) and skin-friction coefficient $c_f$ (bottom). (d)  Inner-scaled mean streamwise velocity $U^+$(top) and Reynolds-stress $\overline{uv}^+$ (bottom) profiles at $Re_{\theta} = 1623$, 2138, and 2588. Dashed and solid lines represent, respectively, the PINN predictions and the reference data. Darker color shows higher $Re_{\theta}$.}
    \label{fig4}
\end{figure*}

\section{Results}

We employ PINNs for solving the RANS equations for four turbulent flow cases, \emph{i.e.}, zero-pressure-gradient (ZPG) boundary layer \citep{EITELAMOR_2014}, adverse-pressure-gradient (APG) boundary layer \citep{bobke_2017}, and turbulent flows over a NACA4412 airfoil \citep{VINUESA2018} and the periodic hill. We also show the applicability of PINNs for the simulation of laminar boundary flows. To evaluate the accuracy of the predictions, we consider the relative $\ell_2$-norm of errors $E_i$ on all the computational points and for the $i^{th}$ variable as:
\begin{equation}
    E_i = \dfrac{||\mathbf{U}_i - \tilde{\mathbf{U}}_i||_2}{||\mathbf{U}_i||_2} \times 100,
    \label{eq:error}
\end{equation}
where $||\cdot||_2$ denotes $\ell_2$-norm, and $\mathbf{U}$ and $\tilde{\mathbf{U}}$ indicate the vectors of reference data and PINN predictions, respectively. Results for $E$ is reported in \Cref{tab:errors} for all the test cases. The dash symbol in this table denotes that the corresponding variable is not calculated in that test case. We discuss each test case in detail in the following.

\begin{table}[h]
    \centering
    \caption{Relative $\ell_2$-norm of errors, as defined in \cref{eq:error}, in the PINNs predictions with respect to the reference data for all the test cases. The dash symbol in the table denotes that the corresponding variable is not calculated in that test case.}
    \vspace{9pt}
    \resizebox{\columnwidth}{!}{%
    \begin{tabular}{lcccccc}
        \hline 
        \vspace{5pt}
        
        \rule{0pt}{3ex} Test & $E_{U}$ & $E_{V}$ & $E_{P}$ & $E_{\overline{u^{2}}}$ & $E_{\overline{uv}}$ & $E_{\overline{v^{2}}}$ \\
        \hline
        
        \rule{0pt}{3ex}FSBL & 0.07 & 0.12 & 0.001 & -- & -- & -- \\
        ZPG & 1.02 & 4.25 & -- & -- & 6.46 & -- \\
        APG & 0.28 &  1.57 & 4.60 & -- & 7.96 & -- \\
        NACA4412 & 1.56 & 2.17 & 7.30 & 9.43 & 11.36 & 4.69 \\
        Periodic hill & 2.77 & 19.70 & 8.61 & 28.18 & 16.70 & 20.24 \\
    \hline
    \end{tabular}
    }
    \label{tab:errors}
\end{table}

\subsection{Falkner–Skan boundary layer (FSBL)}
As the first test case, we solve two-dimensional Navier--Stokes equations for the Falkner--Skan boundary layer at $Re = 100$ using PINNs to show the applicability of PINNs for the laminar boundary layer flows. We consider a boundary layer with adverse-pressure-gradient with $m = -0.08$ leading to $\beta_{\mathrm{\,FS}} = 2m/(m+1) = -0.1739$ where $\beta_{\mathrm{\,FS}} = -0.1988$ corresponds to separation. The inputs of the PINN model are $x$ and $y$ and the outputs are the velocity components and the pressure. We only use the data on the domain boundaries for the velocity components for supervised learning. The reference data is obtained from the analytical solution. The FNN comprises 8 hidden layers, each containing 20 neurons with hyperbolic tangent function (tanh) as the activation function. We first use the Adam optimizer \citep{kingma2017} for the training of the model and then apply Broyden--Fletcher--Goldfarb--Shanno (BFGS) algorithm to obtain a more accurate solution. The optimization process of the BFGS algorithm is stopped automatically based on the increment tolerance. Similar model architecture and training procedures are implemented for the simulation of other test cases using PINNs. 

Results are illustrated in \Cref{fig2} as the contours of $U$, $V$, and $P$ obtained from PINN predictions (left) and reference data (right). The relative $\ell_2$-norm of errors are reported in \Cref{tab:errors}. Our results suggest that excellent predictions can be obtained using PINNs for laminar boundary layer flows. It can be seen in \Cref{fig2} that although we only use velocity components on the domain boundaries as the training data, the PINN model provides accurate predictions for the pressure.
\subsection{ZPG turbulent boundary layer}
For the ZPG boundary layer we employ the simulation data of \cite{EITELAMOR_2014} for a domain range of $1,000 <Re_{\theta} < 7,000$, where $Re_{\theta} = \theta U_{\infty} / \nu$ represents the Reynolds number based on the momentum thickness $\theta$, the free-stream velocity $U_{\infty}$, and the kinematic viscosity $\nu$. For this test case, we consider continuity and streamwise momentum equations as the governing equations and the mean streamwise $U$ and wall-normal $V$ velocity components and the shear Reynolds-stress $\overline{uv}$ as the outputs of the model. 

Results of the PINN simulation are depicted in \Cref{fig3}(a $\sim$ c) in comparison with the reference data.~\Cref{fig3}(a) shows contours of $U$, $V$, and $\overline{uv}$. The relative $\ell_2$-norm of errors for $U$, $V$, and $\overline{uv}$ are 1.02\%, 4.25\%, and 6.46\%, respectively, as reported in \Cref{tab:errors}. The characteristics of the boundary layer are also quantified in terms of the shape factor, defined as the ratio of displacement and momentum thickness, $H_{12} = \delta^* / \theta$, and the skin-friction coefficient $c_f = 2(u_{\tau}/U_{\infty})^2$, where $u_{\tau} = \sqrt{\tau_w / \rho}$ represents the  friction velocity ($\tau_w$ is the mean wall-shear stress and $\rho$ is the fluid density). The relevant velocity and length scales close to the wall are given by $u_\tau$ and $\ell^* = \nu / u_\tau$. The inner-scaled quantities are thus written as, \emph{e.g.}, $U^+ = U/u_{\tau}$ and $y^+ = y/\ell^*$.~\Cref{fig3}(b) shows the shape factor $H_{12}$ and the skin-friction coefficient $cf$ obtained from PINN predictions against the reference data. Moreover,~\Cref{fig3}(c) depicts inner-scaled mean streamwise velocity $U^+$ and Reynolds-stress $\overline{uv}^+$ profiles at three streamwise locations ($Re_{\theta} = 2500$, 4000, and 5500) indicated by white dashed lines in \Cref{fig3}(a) (top left). Our results show excellent accuracy of the PINN simulation.

\subsection{APG turbulent boundary layer}
As the next test case, we simulate the APG turbulent boundary layer for a Reynolds number range of $910<Re_{\theta}<3360$ and at a constant Clauser pressure-gradient parameter $\beta = \delta^* / \tau_w \mathrm{d}P_{\infty} / \mathrm{d}x \simeq 1$ where $P_{\infty}$ is the free stream pressure \citep{bobke_2017}. For this test case, we consider a PINN model with 8 hidden layers, each containing 20 neurons, and continuity and streamwise and wall-normal momentum equations as the governing equations. We observe that, even in the presence of adverse pressure gradient, excellent predictions can be obtained using PINNs as it is shown in \Cref{fig4} where we compared the predictions with the reference data. It can be seen in \Cref{fig4}(a) that although we do not use any turbulence model, the predictions for the shear Reynolds stress $\overline{uv}$ coincide with the reference data. The relative $\ell_2$-norm of errors are reported in \Cref{tab:errors} where the lowest and the highest errors are related to $U$ and $\overline{uv}$ and are equal to 0.28\% and 7.96\%, respectively. \Cref{fig4}(b) shows the accuracy of the PINN predictions against the reference data in terms of boundary layer characteristics, i.e., $H_{\mathrm{12}}$ (top) and $c_f$ (bottom). Moreover, \Cref{fig4}(c) depicts the inner-scaled streamwise velocity (top) and Reynolds-stress (bottom) obtained from the PINN predictions and the reference data at three different $Re_{\theta}$ of 1623, 2138, and 2588. Our results show that the PINN model can provide accurate predictions for the APG turbulent boundary layer.

\subsection{NACA4412 airfoil}
\begin{figure}[th]
    \centering
    \begin{overpic}[width=1\columnwidth]{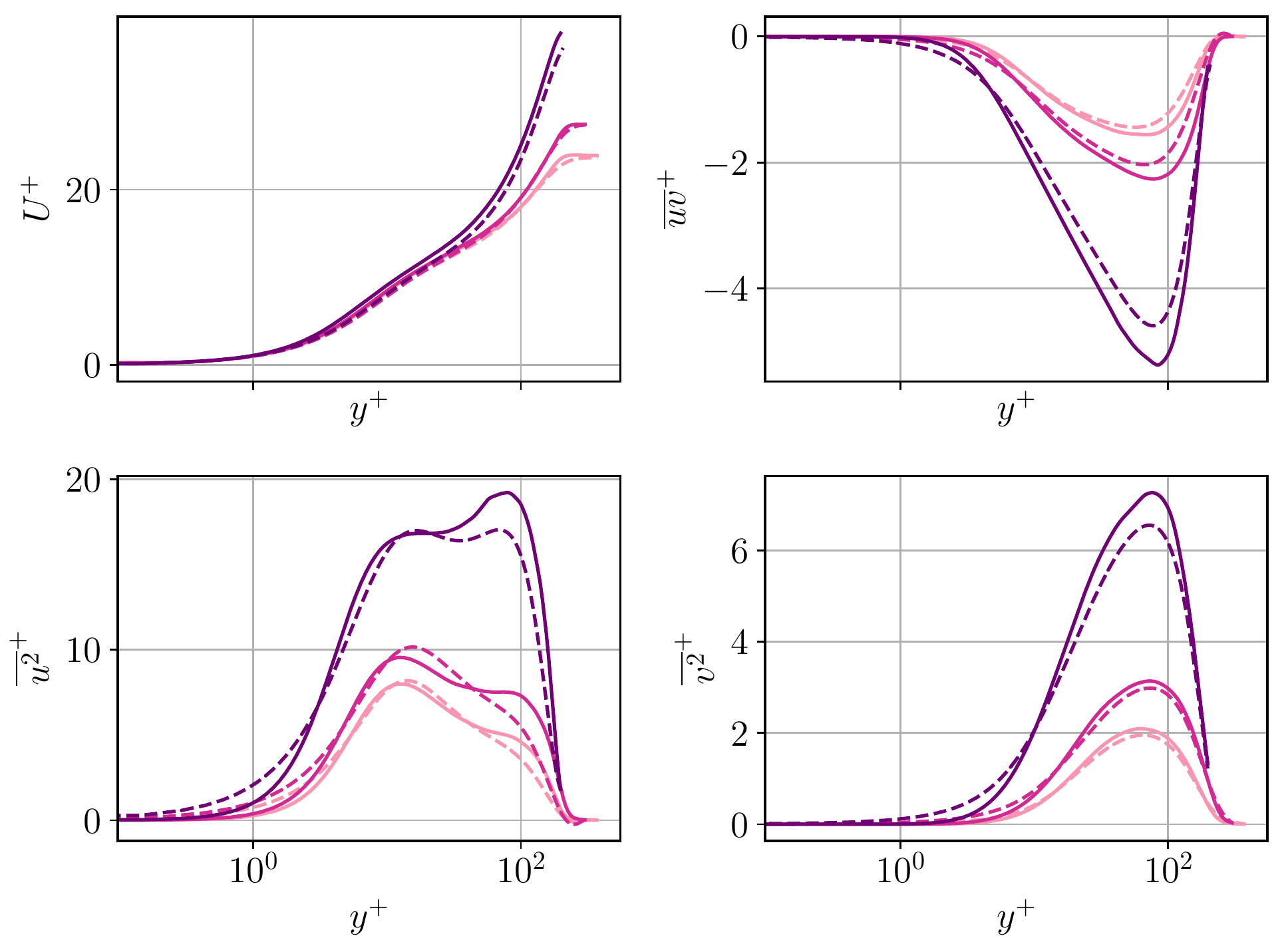}
        \put (-2,73) {(a)}
        \put (50,73) {(b)}
        \put (-2,39) {(c)}
        \put (50,39) {(d)}
    \end{overpic}
    \caption{Simulation results of turbulent boundary layer over the suction side of a NACA4412 airfoil at $Re_c = 200,000$ using PINNs in comparison with the reference data at $x/c \simeq 0.625$, 0.75, and 0.875. (a) Inner-scaled streamwise velocity profile $U^+$. (b, c, d) Profiles of the inner-scaled Reynolds-stress components $\overline{uv}^+$, $\overline{u^2}^+$, and $\overline{v^2}^+$, respectively. Dashed and solid lines represent, respectively, the PINN predictions and the reference data. Darker color shows higher $x/c$.}
    \label{fig5}
\end{figure}

Next, we use PINNs for simulation of the turbulent boundary layer developing on the suction side of a NACA4412 airfoil at the Reynolds number based on $U_{\infty}$ and chord length $c$ of $Re_c = U_{\infty}c/\nu = 200,000$. The data for the training and testing of the model is employed from the work by \cite{VINUESA2018} where results were obtained based on well-resolved large-eddy simulations (LESs) using a spectral-element method. We consider a domain range of $0.5 < x/c < 1$ and a PINN model with 8 hidden layers, each with 20 neurons. Two-dimensional RANS equations are considered as the governing equations. For this test case, we use the wall-normal based spatial coordinates $x_n$ and $y_n$ as the inputs of the FNN and $U$, $V$, $P$, $\overline{u^2}$, $\overline{uv}$, and $\overline{v^2}$ as the outputs. Results are reported as the profiles of the inner-scaled streamwise velocity and Reynolds-stress components at $x/c \simeq 0.625$, 0.75, and 0.875 in \Cref{fig5}. Our results show an excellent agreement between the PINN predictions and the reference data both for velocity and Reynolds-stress components. \Cref{tab:errors} shows the relative $\ell_2$-norm of errors in the PINN predictions. It can be seen that for mean-velocity components $U$ and $V$, the error is less than 3\%. The highest error is related to $\overline{uv}$, and it is equal to 11.36\%.

\begin{figure}[th]
    \centering
    \begin{overpic}[width=1\columnwidth]{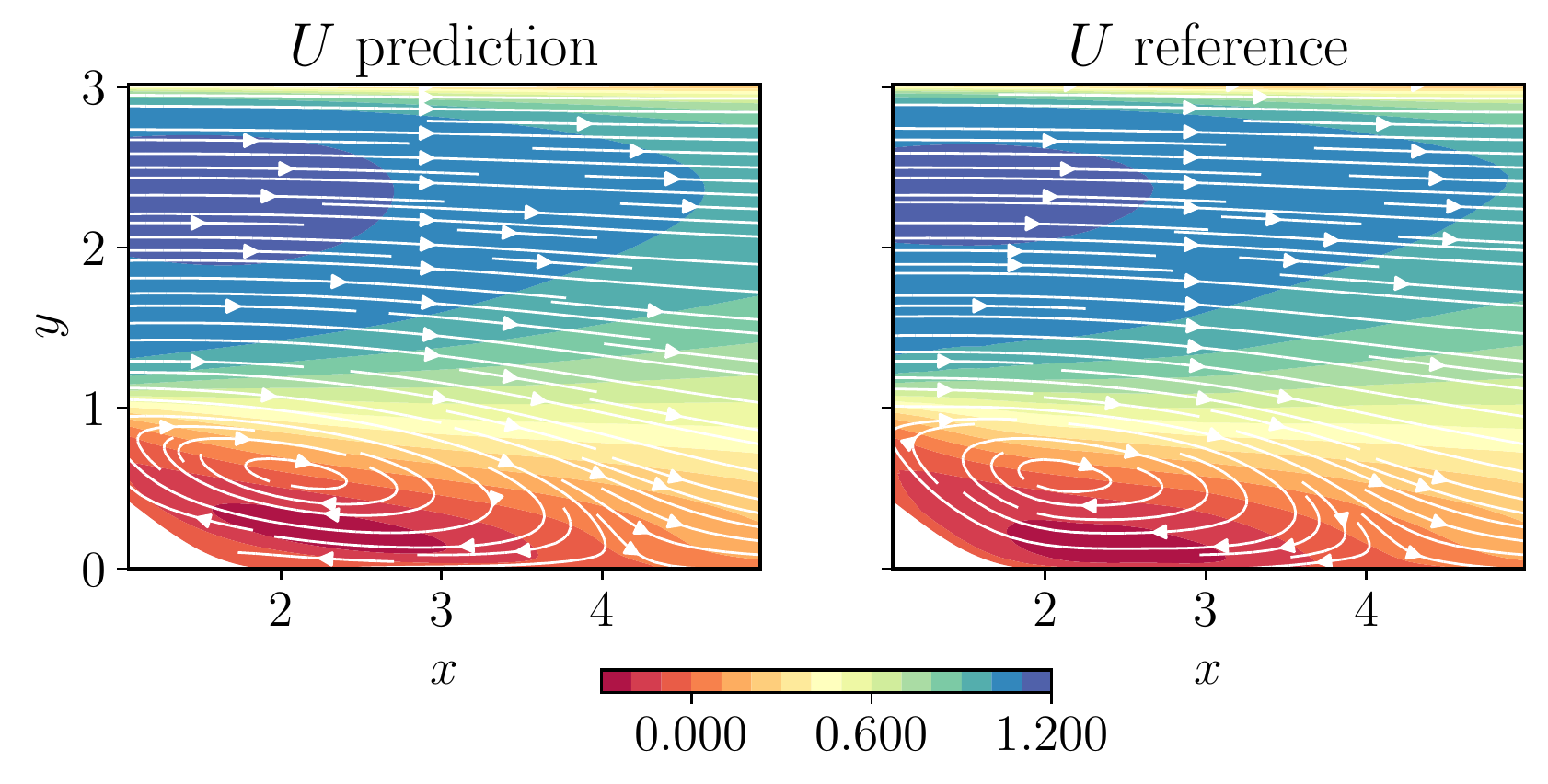}
        \put (-2,47) {(a)}
    \end{overpic}
    \begin{overpic}[width=1\columnwidth]{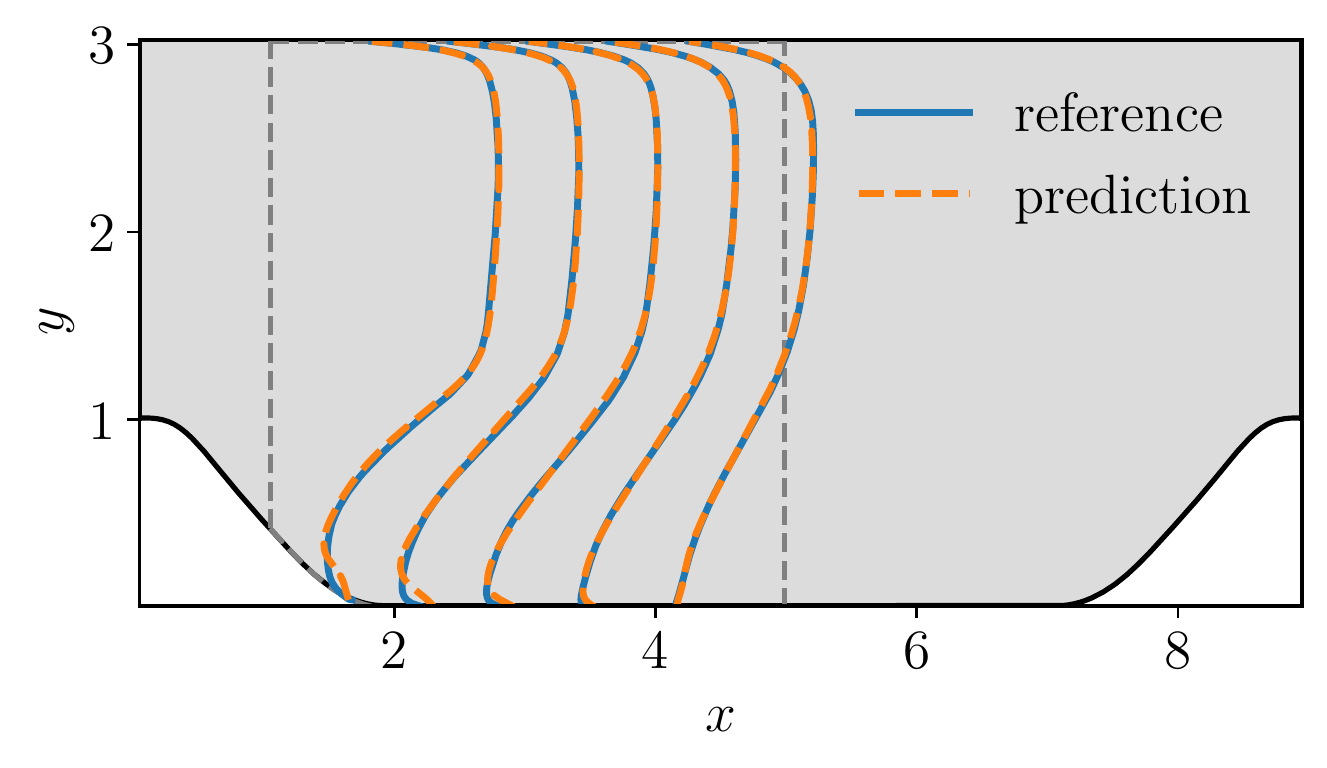}
        \put (-2,55) {(b)}
    \end{overpic}
    \caption{Simulation results of the turbulent flow over a periodic hill at $Re_b = 2,800$ using PINNs in comparison with the reference data. (a) Contour of $U$ and flow streamlines for the PINN simulation (left) and the reference data (right). (b) $U$ profiles at five different streamwise locations.}
    \label{fig6}
\end{figure}

\subsection{Periodic hill}

At last, we evaluate the performance of our proposed framework for solving RANS equations using PINNs in the simulation of the turbulent flow over a periodic hill at the Reynolds number $Re_b = 2,800$ based on the crest height $H$ and the bulk velocity $U_b$ at the crest. The training and testing data are obtained from DNS simulation using a spectral-element method. For this test case, we consider a domain range of $ 1<x/H<5$ as it is depicted in \Cref{fig6} by the grey dashed lines. Similar to the previous test cases, the data on the domain boundaries are used for the training of a PINN model with 8 hidden layers, each containing 20 neurons where two-dimensional RANS equations are considered as the governing equations.~\Cref{fig6}(a) shows the streamwise velocity $U$ contour and the flow streamlines for the PINN predictions (left) and the reference data (right). It can be observed that the PINN model is able to simulate the separated flow without a turbulence model and only by using the data on the domain boundaries and the RANS equations as the governing equations. It should be noted that the velocity and Reynolds-stress components on the top and bottom boundaries are equal to zero due to the no-slip condition. Therefore, we only need the data at the input and output boundaries and pressure on all the boundaries for the supervised learning.~\Cref{fig6}(b) illustrates the streamwise velocity profiles at five different streamwise locations for the PINN predictions and the reference data. Our results show an excellent agreement between the PINN predictions and the reference data. It can be seen in \Cref{fig6} that the extent of the separated region and the reattachment point are accurately predicted by the PINN model. The relative $\ell_2$-norm of errors are reported in \Cref{tab:errors}. The lowest error is related to $U$, and it is equal to 2.77\%. The highest error is equal to 28.18\% and is associated with $\overline{u^2}$.

\section{Conclusions}
We introduced an alternative approach based on PINNs for solving the RANS equations. In contrast to traditional methods, we solve the RANS equations for incompressible turbulent flows without any specific model or assumption for turbulence and through the use of the data on the domain boundaries (including Reynolds-stress components) along with the governing equations that guide the learning process towards the solution. We simulate the Falkner--Skan boundary layer with adverse pressure gradient using PINNs to show the applicability of the model for laminar boundary layer flows. In this case, we only used the data on the boundaries for the velocity components as the training data. Our results suggest the applicability of PINNs for laminar boundary layer flows where excellent predictions can be obtained, even for the pressure. Moreover, we applied PINNs for the simulation of ZPG and APG turbulent boundary layers, and turbulent flows over the NACA4412 airfoil and the periodic hill. We only used the data on the domain boundaries, including Reynolds-stress components, for supervised learning while we considered the residual of the RANS equations as the loss for unsupervised learning. A set of points inside the domain and the points on the domain boundaries are employed to evaluate the residual of the governing equations. From these points we predict the mean-flow quantities. For the points on the boundaries we calculate the supervised loss by comparing the predictions with the training data while for all the points we compute the residuals by constructing the RANS equations using AD. For ZPG and APG boundary layers, we obtained excellent predictions using PINNs where the average relative $\ell_2$-norm of errors are equal to 3.91\% and 3.60\%, respectively. Our results for the NACA4412 airfoil and the periodic hill show that PINNs can provide accurate predictions for the streamwise velocity with less than 3\% error while leading to good accuracy even in the simulation of Reynolds-stress components.
%%%%%%%%%% Insert here acknowledgments if necessary

\Acknowledgments
RV acknowledges the Göran Gustafsson foundation for supporting this research. HE acknowledges the support of the University of Tehran. 
% The investigations presented in this paper have been obtained within the European research project XYZ. 

%%%%%%%%%%%%%%%%%%%%%%%%%%%%%%%%%%%%%%%%%%%%%%%%%%%%%%%%%%%%%%%%%%%%%%%
%%%%%%%%%%%%%%%%%%%%%%%%%%%%%%%%%%%%%%%%%%%%%%%%%%%%%%%%%%%%%%%%%%%%%%%
%
\par \vspace{6pt} \noindent \textbf{\large References}

\renewcommand{\bibsection}{}
\bibliographystyle{apsrev}
\bibliography{sample}

\end{document}